\documentclass[showpacs,preprint]{revtex4}

\usepackage{epsfig}

\usepackage{amsfonts}
\usepackage{color}

\begin{document}

\title{Real stabilization of resonance
states
employing two parameters: basis set size and coordinate scaling }
\author{Federico M. Pont }
\email{pont@famaf.unc.edu.ar}
\author{Pablo Serra}
\email{serra@famaf.unc.edu.ar}
\author{Omar Osenda}
\email{osenda@famaf.unc.edu.ar}

\affiliation{Facultad de Matem\'atica, Astronom\'{\i}a y F\'{\i}sica,
Universidad Nacional de C\'ordoba, C\'ordoba, Argentina and IFEG-CONICET,
Ciudad Universitaria,
X5016LAE C\'ordoba, Argentina}

\begin{abstract}
The resonance states of one- and two-particle Hamiltonians are studied 
using   variational expansions with real basis-set functions. 
The resonance energies,
$E_r$, and widths, $\Gamma$, are calculated using the
density of states and an ${\mathcal L}^2$ golden rule-like formula.
 We present a recipe to select adequately some solutions of the
variational
problem. The set of approximate energies obtained shows
a very regular behaviour with the basis-set size, $N$. Indeed, these particular
variational eigenvalues show a quite simple scaling behaviour and convergence
when $N\rightarrow \infty$. Following the same prescription to choose particular
solutions of the variational problem we obtain a set of approximate
widths. Using the scaling function that characterizes the behaviour of
the approximate energies as a guide, it is possible to find a very good
approximation to the actual value of the resonance width.
\end{abstract}
\date{\today}

\maketitle
\section{Introduction }

The methods used to calculate the energy and the lifetime of a resonance state 
are numerous
\cite{moisereport,reinhardt1996,sajeev2008,bylicki2005,dubau1998,
kruppa1999,kar2004} and, in some cases, has been put forward over strong
foundations~\cite{simon78}.
However, the analysis of the numerical results of a particular method  when
applied to a given problem is far from direct. The complex scaling (complex
dilatation) method \cite{moisereport}, when the Hamiltonian $H$ allows
its use, reveals a resonance state by the appearance of an isolated complex
eigenvalue on the spectrum of the non-Hermitian complex scaled Hamiltonian,
$H(\theta)$ \cite{moisereport}. Of course in an actual implementation the
rotation angle $\theta$ must be large enough to rotate the continuum part of the
spectrum beyond the resonance's complex eigenvalue. Moreover, since most
calculations are performed using finite variational expansions it is necessary
to study the numerical data to decide which result is the most accurate. To
worsen things the variational basis sets usually depend on one (or more)
non-linear parameter. For bound states the non-linear parameter is chosen in
order to
obtain the lowest variational eigenvalue. For resonance states things are not
so simple since they are embedded in the continuum. The complex virial theorem
together with some graphical methods \cite{moiseyev1981,moiseyev1980} allows to
pick the best numerical solution of a given problem, which
corresponds to the stabilized points in the $\theta$ trajectories
\cite{moisereport,moiseyev1981,moiseyev1980}.

Other methods to calculate the energy and lifetime of the resonance, based on
the numerical solution of complex Hamiltonians, also have to deal with the
problem of which solutions (complex eigenvalues) are physically acceptable. For
example, the popular complex absorbing potential method, which in many cases is
easier to implement than the complex scaling method, produces the appearance of
nonphysical complex energy stabilized points that must be removed in order to
obtain only the physical resonances \cite{sajeev2009}.

The aforementioned issues explain, at some extent, why the methods based only
in the use of real ${\mathcal L}^2$ variational functions are often preferred to
analyze resonance states. 
These  techniques reduce the problem to the
calculation of eigenvalues of real symmetric matrices
\cite{kar2004,pont2010,pont2010b,kar2007,kar2007b,kruppa1999}.
Of course, these methods also have its own drawbacks.  One of the main problems
was recognized very early on (see, for example, the work by Hol$\phi$ien and
Midtdal \cite{holfien1966}): if the energy of an autoionizing state is obtained
as an eigenvalue of a finite Hamiltonian matrix, which are the convergence 
properties
of these eigenvalues that lie in the continuum  when the size of the 
Hamiltonian matrix changes?
But in order to obtain resonance-state energies it is possible to
focus the analysis in a global property of the variational spectrum:
the density of states (DOS)\cite{mandelshtam1993}, being unnecessary
 to answer this question.

The availability of the DOS allows to obtain the energy
and lifetime of the resonance in a simple way, both quantities are obtained as
least square fitting parameters, see for example \cite{kar2007,kruppa1999}.
Despite its simplicity, the determination of the resonance's energy and
width based in the DOS is far from complete. There is no  a single  
procedure to asses both,
the accuracy of the numerical findings and its convergence properties, or which
values to pick between the several ``candidates'' that the method offers
\cite{kar2004}.

Recently, Pont {\em et al} \cite{pont2010b} have used {\em finite size scaling}
arguments \cite{ks2003} to analyze the properties of the DOS when the size of 
the Hamiltonian changes. They presented numerical evidence about the critical
behavior of the density of states in the region where a given Hamiltonian has
resonances.
The critical behavior was signaled by a strong dependence of some features of
the density of states with the basis-set size used to calculate it. The
resonance energy and lifetime were obtained using the scaling properties of the
density of states. However, the feasibility of the method to calculate the
resonance lifetime laid on the availability of a known value of the lifetime,
making the whole method dependent on results not provided by itself.

The DOS method relies on the possibility to calculate 
the Ritz-variational eigenfunctions and eigenvalues for
many
different values of the non-linear parameter $\eta$ (see Kar and Ho~
\cite{kar2004}). For each basis-set size, $N$, used, there are $N$ variational
eigenvalues $E_N^m(\eta);\;m=1,\ldots,N$. Each one of these eigenvalues
can be used, at least in principle, to compute a DOS, $\rho_N^m(E)$, resulting,
each one of these DOS in an approximate
value for the energy, $E_r\sim E_N^m$, and width, $\Gamma\sim \Gamma_N^m$, of the
resonance state of
the problem. If the variational problem is
solved for many different basis-set sizes,
there is not a clear cut criterion to pick the ``better'' result
from the plethora of possible values obtained.
 This issue will be addressed in Section~\ref{model}.

In this work, in order to obtain resonance energies and lifetimes, 
we  calculate all the eigenvalues for different basis-set sizes,
and  present a recipe to select adequately certain values of $N$, and 
one eigenvalue for each $N$ elected, that
is, we get a series of variational eigenvalues
$\{E_{N_i}^{m_i}(\eta)\}$.  

The recipe is based on some properties of the variational spectrum
which are discussed in Section~\ref{some-properties}. The properties seem to be
fairly general, making the implementation of the recipe feasible for
problems with several particles. Actually, 
because we use scaling properties for large values of 
$N$, the applicability of the method for systems with more than three particles
could be restricted because the difficulties to handle large basis sets.

The set of approximate resonance energies, obtained from the density of states
of a series of eigenvalues selected following the recipe, shows a very regular
behaviour with the basis set size. This regular behaviour facilitates the use
of finite size scaling arguments to analyze the results obtained, in particular
the extrapolation of the data when $N\longrightarrow \infty$. The extrapolated
values are the most accurate approximation for the parameters of the
resonance state that we obtain with our method. This is the subject of
Section~\ref{recipe}, where we present results for models of one and two particles.

Following the same prescription to choose particular solutions of the
variational problem we obtain a set of approximate widths in
Section~\ref{golden-rule}. Using the scaling function that characterizes the
behaviour of the approximate energies as a guide, it is possible to find a
very good approximation to the resonance width since, again, the data generated
using our prescription seems to converge when
$N\rightarrow \infty$. Finally, in Section~\ref{discusion} we summarize and
discuss our results.

\section{Some properties of the variational spectrum}
\label{some-properties}

When one is dealing with the variational spectrum in the continuum region, some
of its properties are not exploited to obtain more information about the
presence of resonances, usually the focus of interest is the stabilization of
the individual eigenvalues. The stabilization is achieved varying some
non-linear variational parameter. If $\eta$ is the inverse characteristic decaying
length of the variational basis functions, then the spectrum of the kinetic
energy scales as $\eta^2$, moreover, for potentials that decay fast
enough, the spectrum of the whole Hamiltonian {\em also} scales as $\eta^2$ for
large (or small) enough values of $\eta$ (see appendix). 
This is so, since the variational eigenfunctions
are ${\mathcal L}^2$ approximations to  plane waves {\em except} when $\eta$ belongs to
the stabilization region. When $\eta$ belong to the stabilization region of a
given variational eigenvalue, say $E_N^m(\eta)$, then $E_N^m(\eta)\sim E_r$
(where
$E_r$ is the resonance energy) and the variational eigenfunction
$\psi_N^m(\eta)$
has the  localization length of the potential well. We intend to take advantage
of the changes  of the spectrum when $\eta$ goes from small to large enough
values. 

The variational spectrum satisfies the Hylleras-Undheim
theorem or variational theorem: if $N$ is the basis set size, and $E_N^m$ is
the $m-th$ eigenvalue obtained with a variational basis set of size $N$, then
\[
 E_N^m(\eta) \geq  E_{N+1}^m(\eta)  .
\]
Actually, since the threshold of the continuum is an accumulation point, then
for small enough values of $\eta$ and
a given $j\in \mathbb{N}$ there is always a $k \in \mathbb{N}$ such that

\begin{equation}\label{ordering}
 E_N^m(\eta_{small}) >  E_{N+k}^{m+j}(\eta_{small}) .
\end{equation}

For the kinetic energy variational eigenvalues, and for fixed $N,m,j$, and $k$,
if the ordering given by Equation~(\ref{ordering}) holds for some $\eta$ then it
is true for all $\eta$. Of course this is not true for a Hamiltonian with a
non zero potential that support resonance states.
So, we will take advantage of the variational eigenvalues such that for $\eta$
small enough satisfies Equation~(\ref{ordering}) but, for $\eta$ large enough
\begin{equation}\label{reverse}
  E_N^m(\eta_{large}) <  E_{N+k}^{m+j}(\eta_{large}). 
\end{equation}

Despite its simplicity, the arguments above give a complete prescription to
pick a set of eigenstates that are particularly affected by the presence of a
resonance. Choose $N$ and $m$ arbitrary, and then look for the smaller values
of $j$ and $k$ such that the two inequalities, Equations.~(\ref{ordering}) and
(\ref{reverse}) are
fulfilled. So far, all the examples analyzed by us show that if the
inequalities are satisfied for some $j$ and $k$ then they are satisfied too by
the eigenvalues $E_{N+nk}^{m+nj}$, for $n=2,3,4,\ldots$.

\section{Models and methods}
\label{model}

To illustrate how our prescription works we  used two
different model Hamiltonians. The first model, due to Hellmann~\cite{hellman35},
is a one
particle Hamiltonian that models a $N_e$-electron atom. The second one is  a
two particle model that has been used to study the low energy and resonance
states of two electrons confined in a semiconductor quantum
dot~\cite{pont2010b}.

The details of the variational treatment of both models will be kept as concise
as possible. The one particle model has been used before for the determination
of critical nuclear charges for
$N_e$-electron atoms~\cite{sergeev1999}, it also gives reasonable results
for resonance states in atomic anions~\cite{sabre2001} and continuum states
~\cite{colavecchia2009}. The interaction of a
valence electron with the atomic core is modeled by a one-particle
potential with two asymptotic behaviours. The potential behaves correctly in the
regions where electron is far from the atomic core ($N_e-1$ electrons and the
nucleus of charge $Z$) and when it is near the nucleus. The Hamiltonian, in
scaled coordinates $r\rightarrow Z\,r$,  is
\begin{equation}
\label{hamil}
H = -\frac{1}{2} \nabla^2  
-\frac{1}{r} + \frac{\gamma}{r}\left(1-e^{-\delta r}\right) ,
\end{equation}
\noindent where $\gamma=\left(\frac{N_e-1}{Z}\right)$ and $\delta$ is a range
parameter that determines the transition between the asymptotic regimes,   for
distances near the nucleus $V_0(r)\approx-\frac{1}{r}$ and in the case
$r\rightarrow\infty$ the nucleus charge is screened by the $N_e-1$ localized
electrons and $V_{\infty}(r)\approx-\frac{Z-N_e+1}{Zr}$. 

Another advantage of the potential comes from its analytical
properties. In particular this potential is well behaved and
the energy of the resonance states can be calculated using complex scaling
methods. So, besides its simplicity, 
the model potential allows us to obtain the energy of the resonance
by two independent methods and check our results.

The two particle model that we considered describes two electrons interacting
via the Coulomb repulsion and confined by an external potential with spherical
symmetry. We use a short-range potential suitable to apply
the complex
scaling method. The
Hamiltonian $H$ for the system is given by

\begin{equation}
\label{hamiltoniano}
H = -\frac{\hbar^2}{2m} \nabla_{{\mathbf r}_1}^2  
-\frac{\hbar^2}{2m} \nabla_{{\mathbf r}_2}^2  + V(r_1)+V(r_2)+ 
\frac{e^2}{\left|{\mathbf r}_2-{\mathbf r}_1\right|} ,
\end{equation}
where $V(r)=-(V_0/r_0^2)\, \exp{(-r/r_0)}$, ${\mathbf r}_i$ the
position operator of electron $i=1,2$;   $r_0$ and $V_0$
determine  the range and depth of the dot potential.
After re-scaling with $r_0$, in atomic units, the Hamiltonian of 
Equation~(\ref{hamiltoniano}) can be written as
\begin{equation}
\label{hamil2part}
H = -\frac{1}{2} \nabla_{{\mathbf r}_1}^2  
-\frac{1}{2} \nabla_{{\mathbf r}_2}^2 -V_0 e^{-r_1}-V_0
e^{-r_2} + 
\frac{\lambda}{\left|{\mathbf r}_2-{\mathbf r}_1\right|} ,
\end{equation}
where $\lambda=r_0$.

The variational spectrum of the two particle model, Equation~(\ref{hamil2part}),
and all the necessary algebraic details to obtain it, has been studied with
great detail in Reference~\cite{pont2010b} so, until the end of this Section, we
discuss the variational solution of the one particle model given by
Equation~(\ref{hamil}).

The discrete spectrum  and the resonance states of the model given by 
Equation~(\ref{hamil})  can be obtained approximately 
using a real ${\cal L}^2$ truncated basis set $\left\{ \Phi_i(\eta)
\right\}_{1}^{N}$ to construct a $N\times N$ Hamiltonian matrix
$H_{ij}(\eta)=\left\langle \Phi_i(\eta)
\right| H \left| \Phi_j(\eta) 
\right\rangle$. We use the
Rayleigh-Ritz Variational method to obtain the approximations
$\left|\psi_N^m(\eta)\right\rangle$

\begin{equation}\label{variational-functions}
\left|\psi_N^m(\eta)\right\rangle \, =\, 
\sum_{i=1}^N c^{(m)}_{i}(\eta) \left| \Phi_i(\eta) 
\right\rangle \, ,\;\; 
\;\;\;\;m=1,\cdots,N \,.
\end{equation}

For bound states this functions are
variationally optimal. The functions 
$\left| \Phi_i(\eta) \right\rangle$ are

\begin{equation}
 \left| \Phi_i(\eta) \right\rangle = \frac{\eta^{3/2}e^{-\eta r /
2}}{\sqrt{(i+1)(i+2)}}L^{(2)}_{i}(\eta r)
\end{equation}

\noindent and $L^{(2)}_{i}(\eta r)$ are the associated
Laguerre polynomials of $2^{\textrm{nd}}$ order
and degree $i$. The non-linear parameter $\eta$
is used for eigenvalue stabilization in resonance
analysis~\cite{kar2004,kar2007,kar2007b}. Note that $\eta$ plays
a similar role that the finite size of the box in spherical box stabilization
procedures~\cite{cederbaum80}, as stated by Kar {\em et. al.}~\cite{kar2004}.

Resonance states are characterized by isolated complex eigenvalues,
$E_{res}=E_r -i \Gamma/2,\; \Gamma > 0$, whose eigenfunctions are not
square-integrable. These states are considered as quasi-bound states  of
energy $E_r$ and inverse lifetime $\Gamma$. For the Hamiltonian 
Equation~(\ref{hamil}), the resonance energies belong to the positive energy
range \cite{reinhardt1996}.

Using the approximate solutions of Hamiltonian~(\ref{hamil}) we analyze
the DOS method \cite{mandelshtam1993} that has been used extensively to
calculate the energy and lifetime of resonance states, in particular we intend
to show that 1) the DOS method provides a host of approximate values whose
accuracy is hard to assess, and 2) if the DOS method is supplemented by a new
optimization rule, it results in a convergent series of approximate values  for
the energy and lifetime of resonance states.

\subsection*{The DOS method}

The DOS method relies on the possibility to calculate 
the Ritz-variational eigenfunctions and eigenvalues for
many
different values of the non-linear parameter $\eta$ (see Kar and Ho~
\cite{kar2004}).

The localized DOS $\rho(E)$ can be expressed as \cite{mandelshtam1993}
\begin{equation}\label{densidad_sin_suma}
\rho(E)  = \left|\frac{\partial
E(\eta;\gamma,\delta)}{\partial \eta}\right|^{-1} . 
\end{equation}
Since we are dealing with a numerical approximation, we calculate the energies
in a discretization $\left\{\eta_i\right\}$ of the continuous parameter $\eta$.
In this approximation, Equation~(\ref{densidad_sin_suma}) can be written as
\begin{equation}\label{densidad_cal}
\rho_N^m(E) =
\rho(E_N^m(\eta_i;\gamma,\delta))  = \left| 
\frac{E_N^m(\eta_{i+1};\gamma,\delta) -
E_N^m(\eta_{i-1};\gamma,\delta)}{\eta_{i+1} - \eta_{i-1}}\right|^{-1}
\end{equation}

\noindent where $E_N^m(\eta;\gamma,\delta)$ is the $j$-th eigenvalue of the
$N\times N$
matrix Hamiltonian with $\gamma$ and $\delta$ fixed.

In complex scaling methods the Hamiltonian is dilated by a complex factor 
$\alpha=|\alpha|\,e^{-i\theta}$. As was pointed out long time ago by
Moiseyev and coworkers~\cite{moiseyev79}, the role played by $\eta$  and 
$|\alpha|$ are equivalent, in fact, our parameter $\eta$ corresponds to 
$\alpha(\theta=0)$. Besides, the DOS attains its maximum at
optimal values of $\eta$ and $E_r$ that could be obtained with a
self-adjoint Hamiltonian without using complex scaling
methods~\cite{moiseyev1980}.
So, locating the position of the resonance using the maximum of the DOS is
equivalent to the stabilization criterion used in complex dilation methods that
requires the approximate fulfillment of the complex virial
theorem \cite{moiseyev1981}.

The values of $E_r(\gamma,\delta)$ and $\Gamma(\gamma,\delta)$ are
obtained performing a nonlinear fitting of $\rho(E)$, with a Lorentzian
function,

\begin{equation}\label{lorentz}
\rho(E)=\rho_0 + \frac{A}{\pi}\frac{\Gamma/2}{\left[(E-E_r)^2
+(\Gamma/2)^2\right]}.
\end{equation}

One of the drawbacks of this method results
evident: for each pair $\gamma,\delta$ there are several $\rho_N^m(E)$, and
since
each $\rho_N^m(E)$ provides a value for $E_r(N,m)$ and $\Gamma(N,m)$
one has to choose which one is the best.
Kar and Ho \cite{kar2004} solve this problem fitting all the $\rho_N^m(E)$ and
keeping as the best values for $E_r$ and $\Gamma$ the fitting
parameters with the smaller $\chi^2$ value. At least for their data the best
fitting (the smaller $\chi^2$) usually corresponds to the larger $j$. 
This fact has a clear interpretation, if the numerical method approximates
$E_r$ with some $E_r(N,j)$, where $N$ is the basis set size
of the variational method, a large $j$ means that the
numerical
method is able to provide a large number of approximate levels, and so the
continuum of positive-energy states is ``better'' approximated.

\begin{figure}[ht]
\begin{center}
\psfig{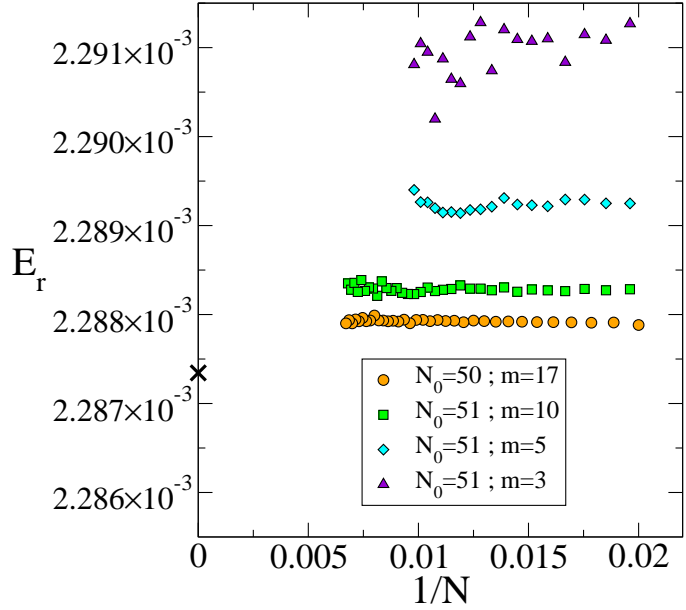}
\end{center}
\caption{\label{prefig1}(color on-line) $E_r(N,m)$ {\em vs} $1/N$ for
$(\gamma=1.125,\delta=0.211)$. The values
were obtained from the peak of the DOS resulting from
Equation~(\ref{densidad_cal}) for several basis set sizes, $N=N_0+3\,k;\;
k=0,\ldots,k_m$. The
(orange) dots data correspond to $N_0=50$
$m=17$ and $k_m=33$; 
the (green) square dots to $N_0=51$,
$m=10$ and $k_m=32$; the (cyan) diamond dots to $N_0=51$
 $m=5$ and $k_m=17$; and the (violet) triangle dots to $N_0=51$
$m=3$ and $k_m=17$. The cross shows the value we obtained using
complex scaling $(\theta=\pi/5,N=70)$.}
\end{figure}

In a previous work \cite{pont2010b} we have shown that a very good approximation
to the energy of the resonance state is obtained considering just the energy
value where $\rho_N^m(E)$ attains its maximum. We denote this value as
$E_r(N,m)$. 
Figure~\ref{prefig1} shows the approximate resonance energy $E_r(N,m)$ for
different basis set size $N$, where $m$ is the index of the variational
eigenvalue used to calculate the DOS.  We used the values
$\gamma=1.125$ and $\delta=0.211$ corresponding to the ones
used before~\cite{sabre2001} in the analysis of $O^{--}$ resonances. The 
Figure~\ref{prefig1}
also shows the value calculated using complex scaling. It is clear that the
accuracy of all the values shown is rather good (all the values shown differ in
less than
6$\times 10^{-6}$), and that larger values on $n$ provide better values for
the resonance energy. These facts are well known from previous works, {\em i.e.}
almost all methods to calculate the energy of the resonance give rather stable
and accurate results for $E_r$. However, the practical importance of this
fact is reduced: these are uncontrolled methods, so the
accuracy of the values obtained 
 from the DOS can not be assessed (without a value independently obtained) and
these values
do not seem to converge to the value
obtained using complex scaling when $N$ is increased and  $n$ is kept fixed.

There is another fact that potentially could render the whole method useless:
for small or even moderate $m$, the values $E_r(N,m)$  become {\em unstable}
(see Figure~\ref{prefig1}) when $N$ is large enough. This last point has been
pointed previously~\cite{holfien1966}.
In
the problem that we are considering is rather easy to obtain a large number of
variational eigenvalues in the interval where the resonances are located,
allowing us to calculate $E_r(N,m)$ up to $m=17$, but this situation is far
from common see, for example, References \cite{pont2010,pont2010b,ferron2009}.

\section{Scaling of the resonance energy}
\label{recipe}

So far we have presented only results about the behaviour of the one particle
Hamiltonian, from now on we will discuss both models, 
Equations~(\ref{hamil}) and
\ref{hamil2part}.

\begin{figure}[ht]
\begin{center}
\psfig{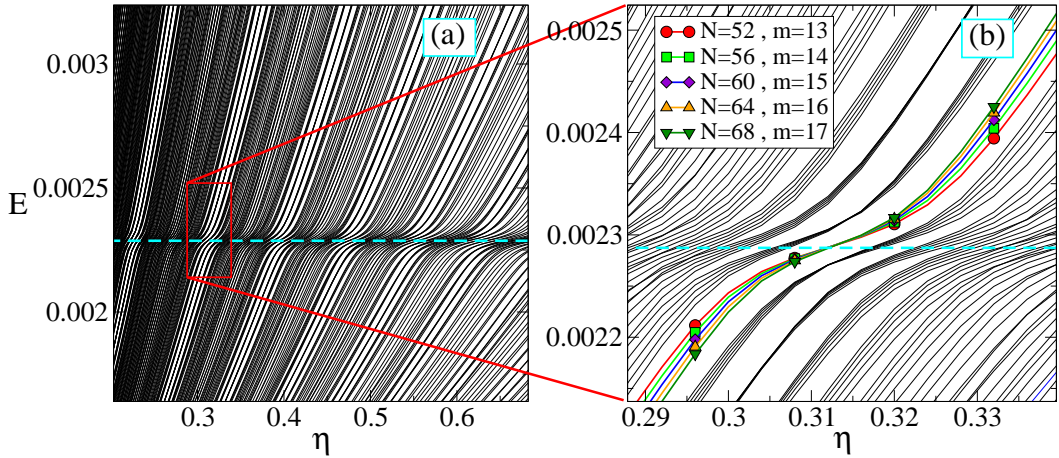}
\end{center}
\caption{\label{prefig2}(color on-line)  (a) Variational spectrum obtained for
several basis set sizes of the  one-particle Hamiltonian. 
There are two noticeable
features associated to changes in the the density of curves. One of
these features corresponds to the stabilization zone, where the derivative of
the eigenvalues is minimal and, correspondingly, the density of states is
larger. The other one corresponds to the gaps in the density of eigenvalues when
$E$ is kept fixed. (b) Detailed view of the data shown
in panel (a). There is a well defined crossing between several
eigenvalues, which form a bundle of states. The bundle is formed by 
$E_{52}^{13}, E_{56}^{14},E_{60}^{15},E_{64}^{16}$, and $E_{68}^{17}$.}
\end{figure}

It is known that the variational eigenvalues $E_N^m(\eta)$ do not present
crossings when they
are calculated for some  fixed values of $N$, {\em i.e} the variational spectrum is
non-degenerate for any finite Hamiltonian matrix. As a matter of fact the
avoided crossings between successive eigenvalues in the variational spectrum are
the watermark of a resonance. An interesting feature emerges when the 
variational
spectrum for many different basis set sizes $N$ are plotted together versus the
parameter $\eta$. Besides the places where $\frac{d E_N^m(\eta)}{d \eta} $
attains its minimum value,
which correspond to the stabilization points, there are some gaps which
correspond to crossings between eigenvalues obtained with different basis set
sizes, see Figure~\ref{prefig2}. Moreover, the crossings corresponds to
eigenvalues with different index $m$, and are the states that
satisfy the inequalities Equations~(\ref{ordering}),and (\ref{reverse}).

\begin{figure}[ht]
\begin{center}
\psfig{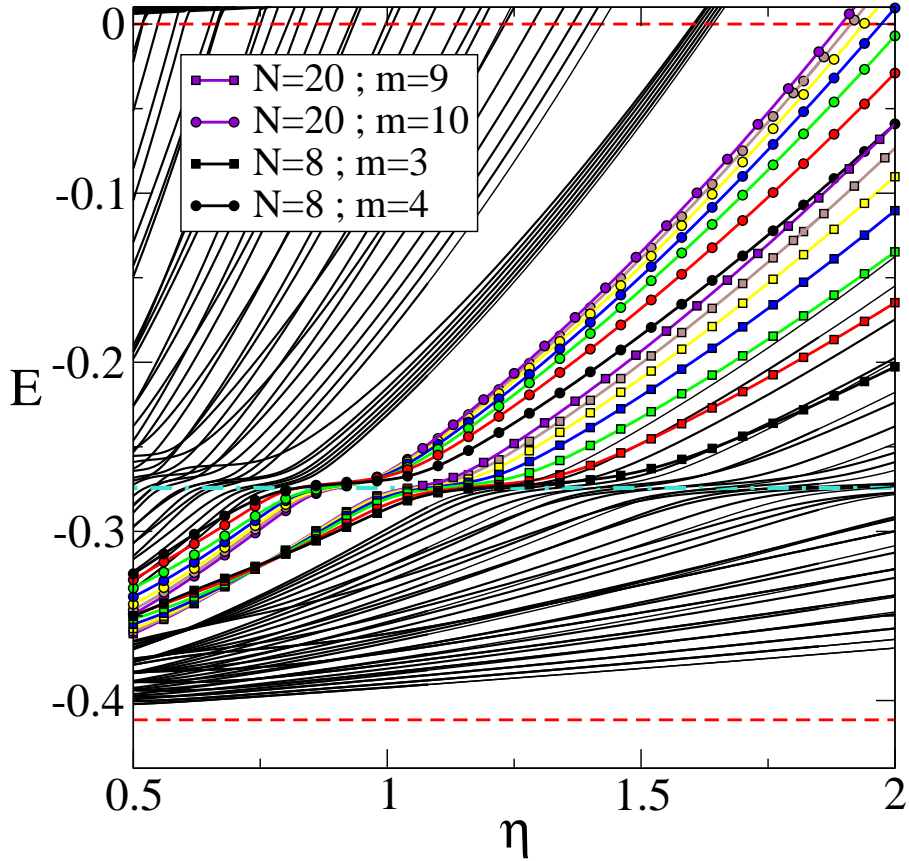}
\end{center}
\caption{\label{bundle2p}(color on-line) Rayleigh-Ritz energies as a
function of the nonlinear parameter $\eta$.
The calculations are for the resonances of the two-electron quantum dot model
used in \cite{pont2010b}.
 The Energies are calculated for $V_0=3$ and $\lambda=1.1$ of equation
(\ref{hamil2part}). The dot and squares denote two different
bundles both calculated with basis-set sizes $N_0=8;\;\Delta N=2$.
The eigenvalue numbers for the dotted bundle are given by $m_0=4;\;\Delta m=1$. 
For the bundle with squares the values are $m_0=3;\; \Delta m=1$.
The eigenvalues corresponding to the minimum ($N=8$) and maximum ($N=20$) 
basis-set size are  single out.
The  dash-dotted cyan line
is the resonance energy calculated using complex scaling. The dashed red lines
indicates the exact energies where the continuum
begins.
}
\end{figure}

It is worth to remark that the main features shown by Figure~\ref{prefig2} are
independent of the number of particles of the Hamiltonian and the particular
values of the threshold of the continuum. Figure~\ref{bundle2p} shows the
behaviour of the variational eigenvalues obtained for the two particle
Hamiltonian Equation~(\ref{hamil2part}). In this case the ionization threshold 
is not the asymptotic
value of the potential, but it is given by the energy of the one particle
ground state. The resonance state came from the two-particle ground state that
becomes unstable and enters into the continuum of states when the quantum dots
becomes ``too small'' to accommodate two electrons. For more details about the
model, see reference~\cite{pont2010b}.

\begin{figure}[ht]
\begin{center}
\psfig{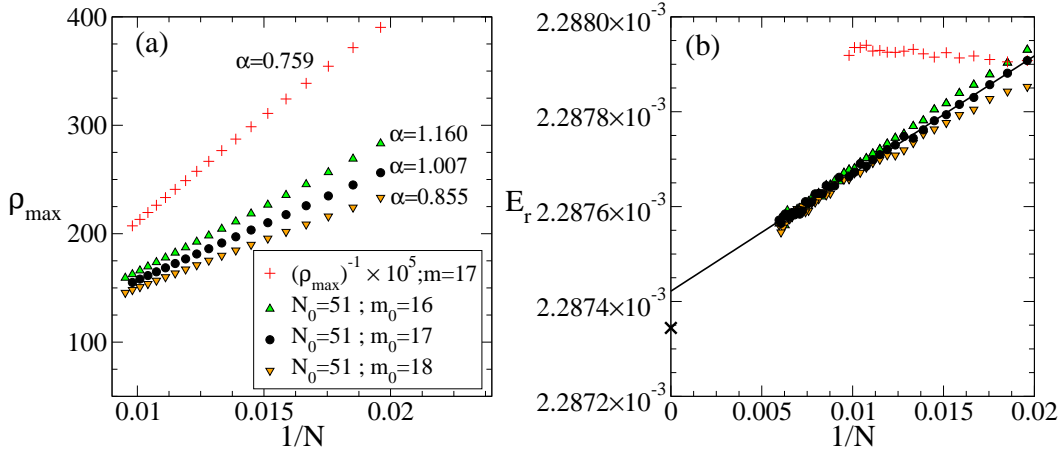}
\end{center}
\caption{\label{prefig3}. {(a) Scaling of the DOS peak {\em vs.} the
inverse basis-set size $N$. One particle Hamiltonian. The figure shows the peak
values with increasing 
$N=51,54,57, \ldots,168\,$,  for $m=17$ fixed  (red crosses). The peaks for a
bundle
of eigenvalues ($N_0=51;\;\Delta N=3;\;m_0=17;\;\Delta m=1$)
show a linear scaling
 (black dots). For series of states that form bundles not associated to a
crossing the
scaling with $1/N$ is greater (
$N_0=51;\,\Delta N=3\,;\,m_0=16\,;\,\Delta m=1$, green triangle up dots)
or lower
($N_0=51;\,\Delta N=3\,;\,m_0=18\,;\,\Delta m=1$, orange triangle down
dots) than $1$. (b) The energy
position of
the DOS peaks as a function of $1/N$. The figure shows that the fixed $m=17$ 
resonance energies (red crosses) do not converge to the resonance  energy 
calculated by  complex scaling (bold black
cross). The scaling for the bundle, and near bundle, series of states is almost
linear and an extrapolated value can be estimated.} }
\end{figure}

 The left panel of 
Figure~\ref{prefig3} shows the behaviour of the maximum value of the DOS,
$\rho_{max}^{-1}(N,m)$, for the one particle Hamiltonian,
obtained for different basis-set sizes and fixed $m$ (in this case
$m=17$), and the $\rho_{max}(N,m)$  obtained choosing a ``bundle'' of
states that are linked by a
crossing, these states have $N=51,54,57,
\ldots,168$ and
$m=16,17,18,\ldots,55$ respectively. From our numerical data, the maximum value
of
the DOS scales with the basis-set size following two different prescriptions.
For $m$ fixed, $\rho_{max}(N,m)\sim N^{\alpha}$, with $\alpha>0$, while
 when the pair $(N,m)$
is chosen from the set of pairs that label a bundle of states
$\rho_{max}(N,m)\sim N^{-\beta}$, with $\beta>0$.
In particular,
for $m=17$ we get that $\rho_{max}(N,17) \sim N^{0.759}$, and $\rho_{max}(N,m)
\sim N^{-1}$ when $(N,m) = (51,17),(54,18),(57,19), \ldots$.

Of course we can pick sets of states that are not related by a crossing. For
instance, we also picked  sets with a simple prescription as follows: 
choose a given
initial pair $(N_0,m_0)$ and form a set of states with the states labeled by
$(N_0,m_0), (N_0+\Delta N,m_0+\Delta m), (N_0+2\Delta N,m_0+2\Delta m)$ and so
on. Figure~\ref{prefig3}(a) shows two examples obtained choosing
$N_0=51$, $\Delta N=3$ and $m_0=16$ and $m_0=18$ both with $\Delta m=1$.  
Quite interestingly, the data in Figure~\ref{prefig3} show that the scaled
maxima of the DOS for a bundle and  two
different sets seem to converge to the {\em same} value when $N\rightarrow
\infty$, but only for the bundle  the scaling function is
$N^{-1}$. The advantage obtained from picking those eigenvalues
$E_N^m(\eta)$ that belong to a given bundle is still more evident when the
corresponding DOS and $E_r$ are calculated. The right panel of
Figure~\ref{prefig3} shows the $E_r$ obtained from the DOS  whose maxima
are shown  in the
left panel. It is rather evident that these values now seem to converge,
besides, the extrapolation to $N\rightarrow \infty$ results in  a more
accurate approximate value for $E_r$. In contradistinction, the values for
$E_r$ corresponding to a fixed index $m$ (the values shown in the Figure 
\ref{prefig3}
correspond to $m=17$) do not seem to converge anywhere close to the
value obtained using complex rotation.

\begin{figure}[ht]
\begin{center}
\psfig{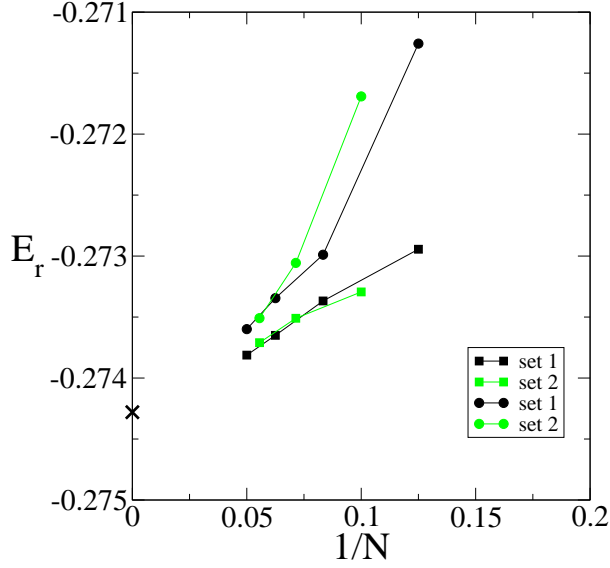}
\end{center}
\caption{\label{er2par}  Resonance Energy for the two bundles 
(dots and squares) showed in
Figure~\ref{bundle2p}.
Since there is an
even-odd behavior the points of each bundle are grouped in two sets.
Data in sets 1 ( Black ) correspond to $N=8,12,16,20$, and for sets 2 
(green) correspond to
$N=10,14,18$. 
The black cross shows the width calculated using complex
scaling. }
\end{figure}

Figure~\ref{er2par} shows the resonance energies obtained from the bundles of
states shown in Figure~\ref{bundle2p} for the two-particle model. Since the
numerical solution of this model is more complicated than the solution of the
one-particle model the number of approximate values is rather reduced. However,
it seems that the data also supports a linear scaling of $E_r(N,m)$ with $1/N$.

\section{Scaling of the resonance width}
\label{golden-rule}

Many  real algebra methods to calculate resonance energies use
a Golden Rule-like formula to calculate the resonance width.  
In this section we will use the formula and
stabilization procedure  proposed by Tucker and Truhlar \cite{tucker1987} that
we will describe briefly for completeness. This projection formula seems
to work better for one-particle models. For two-particle models its utility has
been questioned~\cite{manby1997}, so to analyze the width of the resonance
states
of the quantum dot model we fitted the corresponding DOS using
Equation~(\ref{lorentz}).

The method of Tucker and Truhlar \cite{tucker1987} is implemented by the
following steps.
Choose a basis $\left\{
\phi_j(\eta)\right\}$ where $\eta$ is a non-linear  parameter. 
Diagonalize the Hamiltonian using up to $N$ functions of the basis.
Look for the stabilization value ${\mathcal E}_{res}$ and its corresponding
eigenfunction $\psi_{res}$ which are founded for some value $\eta^r$. Define
the projector 
\begin{equation}
 Q^{\eta} = \left|\psi^{r,\eta}\right\rangle \left\langle \psi^{r,\eta}
\right| ,
\end{equation}
where $ \left|\psi^{r,\eta}\right\rangle$ is the normalized projection of
$\psi_{res}$ onto the basis $\left\{ \phi_j(\eta)\right\}_{j=1}^N$ for any
other $\eta$.

Diagonalize the Hamiltonian $\tilde{H} = (1-Q^{\eta}) H (1-Q^{\eta})$ in
the basis $\left\{ \phi_j(\eta)\right\}_{j=1}^N$, again as a function of
$\eta$, and find a value $\eta^c$ of $\eta$ such that
\begin{equation}\label{seg-estabili}
 E_n(\eta^c) = {\mathcal E}_{res},
\end{equation}
where $E_n(\eta^c)$ denotes eigenvalue $n$ of the projected Hamiltonian for the
scale factor $\eta^c$, and $\chi_n(\eta^c)$ is the corresponding
eigenfunction.

With the previous definitions and quantities, the resonance width $\Gamma$ is
given by
\begin{equation}\label{golden2}
\Gamma = 2\pi \rho(\eta^c, {\mathcal E}_{res}) \left|\right\langle
\psi_{res} \left| H \right| \chi_n(\eta^c) \left\rangle \right|^2 ,
\end{equation}
where
\begin{equation}
\rho(\eta^c, {\mathcal E}_{res}) = 2/\left[E_{n+1}(\eta^c) -
E_{n-1}(\eta^c) \right].
\end{equation}

\begin{figure}[ht]
\begin{center}
\psfig{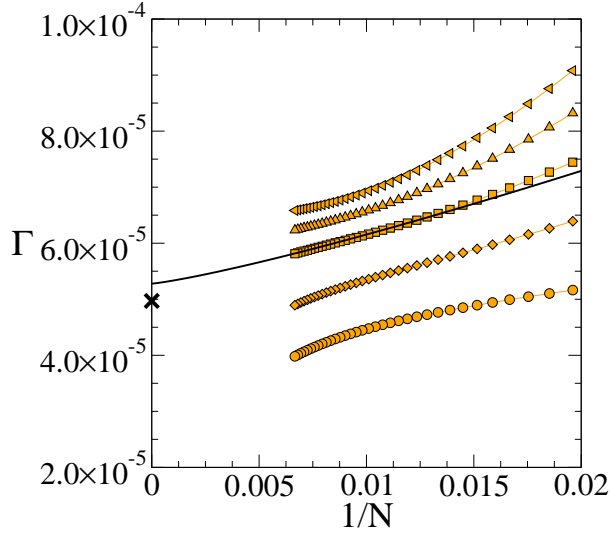}
\end{center}
\caption{\label{prefig4} Scaling of the width vs. the inverse basis-set size
$N$. All points correspond to the same bundle of states $(N=51,54,57,\ldots, 150
\,;\,
m=17,18,19,\ldots,50)$. The curves were obtained with different values of the
projection eigenvalue $n$ ( $1,2,3,4,5$, from bottom to top) . The
black line is a fit for $n=3$ (Equation~(\ref{scaling_gama})) and the lowest $n$
curve
that shows a scaling exponent greater than one~($\gamma(3)\sim 1.2$). }
\end{figure}

Despite some useful insights, the procedure sketched above does not determine
all the intervening quantities, for instance there are many solutions to
equation~(\ref{seg-estabili}) and, of course, the stabilization method provides
several good candidates for $\psi_{res}$ and $\eta^r$.

We are able to avoid some of the indeterminacies associated to the
Tucker and Truhlar procedure using a bundle of states associated to a crossing,
so $\psi_{res}$ and $\eta^r$ are given by any of the eigenfunctions
associated to a bundle and  $\eta^r(N,m)$ comes from the stabilization
procedure. Then we construct projectors 
\begin{equation}
Q^{\eta}_{N,m} = \left|\psi_N^m (\eta)\right\rangle \left\langle
\psi_N^m(\eta)
\right|,
\end{equation}
where $\psi_N^m(\eta)$ is one of the variational eigenstates that belong
to a bundle of states. With the projectors $Q^{\eta}_{N,m}$ we construct
Hamiltonians $\tilde{H}$, and find the solutions to the problem
\begin{equation}\label{tercera-estabili}
E_n(\eta_n^c) = {\mathcal E}_{res}.
\end{equation}
Since there is not an a priori criteria to choose one particular solution of
Equation~(\ref{tercera-estabili}) we show our numerical findings for several
values of $n$. Figure~\ref{prefig4} shows the
behaviour of the resonance width calculated with Equation~(\ref{golden2}), where
we have used $\psi_{51}^{17},\,\psi_{54}^{18},\,\psi_{57}^{19},\ldots$
 as $\psi_{res}$ and $\chi_n$, where $n=1,2,3,4$ and $5$. 

Despite that the different sets corresponding to different values of $n$ do not
converge to any definite value, for $N$ large enough all the sets scale as
$N^{-\gamma(n)}$, with $\gamma(n)>0$. Since the resonance energy scales as
$N^{-1}$, at least when a bundle of states with a crossing is chosen to
calculate approximations (see Figure~\ref{prefig3}), we suggest that the right
scaling for $\Gamma$ is given by $\gamma=1$. Of course for a given basis size,
particular variational functions, stabilization procedures and so on, we can
hardly expect to find a proper set of $\Gamma$ whose scaling law would be
$N^{-1}$. Instead of this we propose that the data in the right panel can be
fitted by 
\begin{equation}\label{scaling_gama}
 \Gamma_{N,n} \sim A(\gamma(n)) + B(\gamma(n)) N^{-\gamma(n)}, \quad \gamma(n)>0
.
\end{equation}
Then the best approximation for the resonance width is obtained
fitting the curve and selecting the $\Gamma_b$ as $A(\gamma(n))$ for
$\gamma(n)$ the closest value to one.

\begin{figure}[ht]
\begin{center}
\psfig{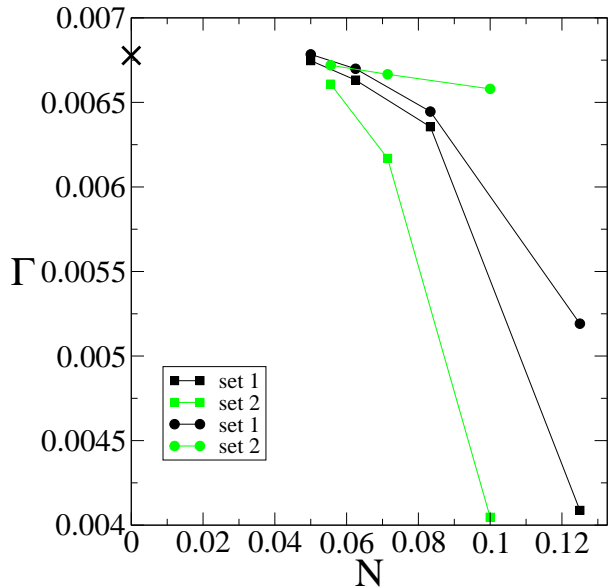}
\end{center} 
\caption{\label{gama2par}(color on-line) Resonance widths for the two bundles
(dots and squares) showed in
Figure~\ref{bundle2p}.
Since there is an
even-odd behavior the points of each bundle are grouped in two sets.
As in figure \ref{er2par}, data in sets 1 ( Black ) correspond to 
$N=8,12,16,20$, and for sets 2
(green) correspond to
$N=10,14,18$.
The widths where calculated using the method described in 
Reference \cite{kar2004}.
The black cross is the width calculated using complex
scaling. }
\end{figure}

As pointed in Reference~\cite{manby1997}, the projection technique to calculate
the width of a resonance can be implemented if a suitable form of the
projection operator can be found. As this procedure is marred by several issues
we used the DOS method to obtain the approximate widths of a resonance state of
the two particle model. Figure~\ref{gama2par} shows the widths calculated
associated to the energies shown in Figure~\ref{er2par}, the parameters of the
Hamiltonian are exactly the same.

There is no obvious scaling function that allows the extrapolation of the data
but, even for moderate values of $N$, it seems as the data converge to the
value obtained using complex scaling.

\section{Conclusions}
\label{discusion}
In this work we analyzed the convergence properties of real ${\cal L}^2$
basis-set methods to obtain resonance energies and lifetimes. 
The convergence of the energy with the basis-set size for bound states is
well understood, the larger the basis set the better the results and
these methods converge to the exact values for the basis-set size going
to infinite (complete basis set).
This idea is frequently applied to resonance states. 
The increase of the 
basis-set size in some commonly used methods does not improve the
accuracy of the value obtained for the resonance energy $E_r$, as showed in 
Figure~\ref{prefig1}.
This undesirable behavior comes from the fact that the procedure is not
variational as in the case of bound states. Moreover, the exact resonance
eigenfunction does not belong to the  Hilbert space expanded by the complete
basis set. In this work we presented a prescription to pick
a set or bundle of states that has linear convergence properties for small width
resonances. This procedure is robust because the choice of different
bundles results in very similar convergence curves and energy values. 
In fact, in the  method described here, the pairs $(N,m)$ of the bundles
play the role of a second stabilization parameter together with the
variational parameter $\eta$.
of a second 
We tested the method in others one and two particle systems and the general 
behavior of them is the same. The results
 are very good in all cases leading to an improvement in the calculation of 
the resonance energies. Nevertheless we have to note that the method could
no be applied in cases where two or more resonance energies lie very close
because the overlapping bundles.

 The lifetime calculation is more subtle. The use of golden-rule-like
formulas,  as we applied here, always give several possible outcomes for the
width $\Gamma_{N,n}$, corresponding to different pseudo-continuum states
$\left|\chi_n\right\rangle $. The projection
technique, Equation~(\ref{golden2}), is not the exception and it is not possible
to
select {\em a priori} which value of $\left\{\Gamma_{N,n}\right\}$ is the most
accurate. The linear convergence
of the DOS with basis-set size suggests that the scaling in the lifetime value,
in accordance with the Energy scaling, should be linear. Regrettably, the
projection method gives discrete sets of values which cannot be tuned to obtain
an exact linear convergence. Our recipe is to choose the
set $\Gamma_{N,n}$ whose scaling
is closest to the linear one, then the best estimation for the resonance
width is obtained from extrapolation.

Many open questions remain on the analysis of the different convergence
properties of resonance energy and lifetime. The method
presented here to obtain the resonance energy from convergence properties works
very well, but the appearance of bundles in the spectrum is not completely
understood.  Even there is not a rigorous
proof, the numerical evidence supports the idea that the behaviour of the
systems studied here is quite general.

\appendix*
\section*{Appendix}

In this appendix we give arguments that support our assumptions on
the scaling of the eigenenergies with the basis-set parameter $\eta$.
We present our argument for  one body Hamiltonians, but it is
straightforward to generalize to more particles with pair interactions
decaying fast enough at large distances.

First et all, we present two  very known results that we need later.

\begin{enumerate}
\item Let  $A$ be an  $n \times n$ matrix with all
its matrix elements
having the form $a_{i,j}(\eta)=a_{i,j}(1) \,f(\eta) $, where $f(1)=1$, then
if $f(\eta)\ne 0$ the eigenvalues of $A$ scales with $f(\eta)$:
\begin{equation}
\label{asa}
det[A(\eta)-\lambda(\eta)\,I]=0\;\Rightarrow
det[A(1)-\frac{\lambda(\eta)}{f(\eta)}\,I]=0 \Rightarrow
\lambda(1)=\frac{\lambda(\eta)}{f(\eta)} \,.
\end{equation}

\item  Let $A, \,B$  $n \times n$ be symmetric matrices with $|b_{i,j}|<const.$,
and $\alpha_i,\,\beta_i,\,\lambda_i,\;i=1,\cdots,n$ the eigenvalues of
$A, \,B$ and $A+\varepsilon B$ respectively,  in nondecreasing order,
then, by the minimax principle \cite{wilkinson65}
\end{enumerate}

\begin{equation}
\alpha_i+\varepsilon  \beta_1 \leq\lambda_i\leq\alpha_i+\varepsilon  \beta_n
\end{equation}

Consider a spherical one-particle potential with compact support,
$v(r)=0$ if $r>R$, and finite, $|v(r)|<\infty\;\forall r$ (both conditions
could be relaxed, but we adopt them for simplicity).
Let the basis-set functions be of the form
$\Phi_n(\eta,r) = c_n(\eta)\, \Phi_n(\eta r)$ with $||\Phi_n(\eta,r)||=1 $
and $c_n(1)=1$, then the coefficients take the form
$ c_n(\eta)=c_n(1) \eta^{3/2}$.

With these assumptions, the matrix elements for the kinetic energy are

\begin{equation}
\begin{array}{l}
T_{m,n}(\eta)\,=\,c^*_m(1)c_n(1) \,\eta^3\,
\int\;d^3x \Phi_m(\eta,r)  \frac{1}{r^2}\,\frac{\partial}{\partial r}\,\left(
r^2 \frac{\partial}{\partial r} \right)  \Phi_n(\eta,r) \,=\,\\
\mbox{}\\
c^*_m(1)c_n(1) \,\eta^2\,
\int\;d^3x\; \Phi_m(1,r)  \frac{1}{r^2}\,\frac{\partial}{\partial r}\,\left(
r^2 \frac{\partial}{\partial r} \right)  \Phi_n(1,r) \,=\,
\eta^2  T_{m,n}(1) \,,
\end{array}
\end{equation}

\noindent and then, by Equation (\ref{asa}), all the eigenvalues of the kinetic
energy have the same scaling with $\eta^2$.
We have to show that, in both
limits, $\eta\rightarrow 0$ and  $\eta\rightarrow \infty$, for all the
potential matrix  elements  hold $v_{m,n}/\eta^2 \ll 1$, and then, by the
Wielandt-Hoffman theorem \cite{wilkinson65}, the eigenenergies are a perturbation
 of  the eigenvalues of the kinetic energy.

\subsubsection*{$\eta \rightarrow 0$}

In this limit  $ \eta \,R \ll 1$
$ \Phi(\eta r)\simeq   \Phi(0)$ for $r\varepsilon (0,R]$, then
\begin{equation}
v_{m,n}\simeq \left[4\,\pi\,\Phi^*_m(1,0)\,\Phi(1,0)\,\int_0^R\,r^2\,dr\,v(r)
\right]\;\eta^3\,=\,\hat{v}_{m,n}\,\eta^3\,,
\end{equation}

\noindent where $\hat{v}_{m,n}$ do not depend on $\eta$. Then, in this limit
 $v_{m,n}/\eta^2 \sim \eta \ll 1$.

\subsubsection*{$\eta \rightarrow \infty$}

For $ \eta \,R \rightarrow \infty$ we obtain for the potential matrix elements

\begin{equation}
\begin{array}{ll}
v_{m,n}\,=\,& c^*_m(1)c_n(1) \,\eta^3\,\int_0^R\,r^2\,dr\, \Phi^*_m(\eta\,r)\,
v(r) \, \Phi(\eta\,r)\,=\, \\
\mbox{}&\mbox{}\\
\mbox{}&c^*_m(1)c_n(1) \,\int_0^{\eta R}\,r^2\,dr\, \Phi^*_m(r)\,v(r/\eta)\,\Phi(r)\,\simeq\\
\mbox{}&\mbox{}\\
\mbox{}&c^*_m(1)c_n(1) \,v(0)\,\int_0^{\eta R}\,r^2\,dr\, \Phi^*_m(r)\,
\Phi(r)\,\stackrel{\longrightarrow}{\eta \rightarrow \infty}\,\\
\mbox{}&\mbox{}\\
\mbox{}&c^*_m(1)c_n(1) \,v(0) \,.
\end{array}
\end{equation}

\noindent Then also in this limit we obtain
$v_{m,n}/\eta^2 \sim 1/\eta^2  \ll 1$.

\acknowledgments We would like to acknowledge SECYT-UNC, CONICET,
and MINCyT C\'ordoba for partial financial support of this project.

\end{document}